\documentclass[12pt,aps,pra,groupedaddress,amssymb,amsfonts,nofootinbib,tightenlines]{revtex4} 
\usepackage{graphicx} 

\graphicspath{{./}{graphics/}}
\usepackage{times,color}
\definecolor{purple}{rgb}{0.625,0.125,0.9375}

\usepackage{ekqc_b}

\ignore{
rm -R /tmp/thresh; mkdir /tmp/thresh
cp `texfls thresh.log | perl -e '$a=<>; $a =~ s:/\S*(revtex4|natbib)\S*(\s|$)::g; print "$a\n"'` /tmp/thresh
find /tmp/thresh -type f -name '*.pdf' -exec pdftops -eps {} \; -exec rm {} \;
find /tmp/thresh -type f -name '*.jpg' -exec perl -e '$f=shift; $g=f$; $g=~s/.jpg/.eps/; exec("convert $f $g");' {} \; -exec rm {} \;
perl -i -n -e 's/^

cp /usr/share/texmf/tex/latex/tools/calc.sty /tmp/thresh/phcalc.sty
perl -i -n -e 's/ProvidesPackage\{calc\}/ProvidesPackage\{phcalc\}/; print;' /tmp/thresh/phcalc.sty
perl -i -n -e 's/usepackage\{calc\}/usepackage\{phcalc\}/; print;' /tmp/thresh/ekqc_b.sty

(cd /tmp/thresh; tar czvf thresh.tar.gz *)
}

\ignore{
rm -R /tmp/thresh; mkdir /tmp/thresh
cp `texfls thresh.log | perl -e '$a=<>; $a =~ s:/\S*(revtex4|natbib)\S*(\s|$)::g; print "$a\n"'` /tmp/thresh
find /tmp/thresh -type f -name '*.pdf' -exec pdftops -eps {} \; -exec rm {} \;
find /tmp/thresh -type f -name '*.jpg' -exec perl -e '$f=shift; $g=f$; $g=~s/.jpg/.eps/; exec("convert $f $g");' {} \; -exec rm {} \;
perl -i -n -e 's/^

cp /usr/share/texmf/tex/latex/tools/calc.sty /tmp/thresh/phcalc.sty
perl -i -n -e 's/ProvidesPackage\{calc\}/ProvidesPackage\{phcalc\}/; print;' /tmp/thresh/phcalc.sty
perl -i -n -e 's/usepackage\{calc\}/usepackage\{phcalc\}/; print;' /tmp/thresh/ekqc_b.sty

cp /usr/share/texmf/tex/latex/base/ifthen.sty /tmp/thresh/phifthen.sty
perl -i -n -e 's/ProvidesPackage\{ifthen\}/ProvidesPackage\{phifthen\}/; print;' /tmp/thresh/phifthen.sty
perl -i -n -e 's/usepackage\{ifthen\}/usepackage\{phifthen\}/; print;' /tmp/thresh/ekqc_b.sty

(cd /tmp/thresh; tar czvf thresh.tar.gz *)
}

\begin{document}
\title{Scalable Quantum Computation in the Presence of Large Detected-Error Rates}
\author{E. Knill}
\email[]{knill@boulder.nist.gov}
\affiliation{National Institute of Standards and Technology} 

\date{\today}
\begin{abstract}
The tolerable erasure error rate for scalable quantum computation is
shown to be at least $0.292$, given standard scalability assumptions.
This bound is obtained by implementing computations with generic
stabilizer code teleportation steps that combine the necessary
operations with error correction.  An interesting consequence of the
technique is that the only errors that affect the maximum tolerable
error rate are storage and Bell measurement errors. If storage errors
are negligible, then any detected Bell measurement error below $1/2$ is
permissible.  Another consequence of the technique is that the
maximum tolerable depolarizing error rate is dominated by how
well one can prepare the required encoded states. For example, if
storage and Bell measurement errors are relatively small, then
independent depolarizing errors with error rate close to $0.1$ per qubit
are tolerable in the prepared states. The implementation overhead is
dominated by the efficiency with which the required encoded states can
be prepared. At present, this efficiency is very low, particularly
for error rates close to the maximum tolerable ones.
\end{abstract}


\maketitle

\section{Introduction}

One of the most significant obstacles to realizing scalable quantum
computation is physical noise that can quickly destroy the information
contained in the computational state. It is now known that, provided
the physical noise is sufficiently weak and local in space and time,
scalable quantum computation is possible by means of fault-tolerant
encodings of quantum
information~\cite{shor:qc1996a,kitaev:qc1996a,knill:qc1996b,aharonov:qc1996a,aharonov:qc1999a,knill:qc1997a,knill:qc1998a,gottesman:qc1997a,preskill:qc1998a}.
Therefore, in studying a proposed physical implementation, a key
question is whether the noise in the implementation is sufficiently
low for scalability to be possible in principle. More importantly, is
it feasible in practice? Answers to these questions depend
significantly on the specific noise in the implementation, as well as
on the way in which the quantum operations necessary for computation
are realized. Nevertheless, the current consensus is that the error
should be below an error rate of $10^{-4}$ per
operation~\cite{gottesman:qc2000b,hughes:qc2002a}.  There have been
numerous suggestions that $10^{-4}$ is a pessimistic estimate of the
error threshold (below which scalable quantum computation is possible)
and certainly does not apply uniformly to all needed quantum
operations~\cite{zalka:qc1996a,aharonov:qc2002a,duer:qc2002a,steane:qc2002a}.
Steane~\cite{steane:qc2002a} makes it clear that thresholds are at
least an order of magnitude higher under reasonable assumptions.

If the nature of the errors is constrained, then the maximum tolerable
error rate can be much higher.  A notable example of this is in
efficient linear optics quantum computation
(eLOQC~\cite{knill:qc2000e}) where, by design, the errors are
dominated by unintentional but detected measurements of
$\sigma_z$. For this error model, any error rate below $0.5$ is
tolerable~\cite{knill:qc2000a,knill:qc2001e,note1}.  That the
tolerable error rate is so high is due to the great advantages of
being able to detect errors before attempting correction. Here it is
shown that these advantages also apply to the erasure error model.  In
this model, errors are detected but otherwise unknown.  Another way of
thinking about this error model is to imagine that the only error is
loss of qubits, and whether or not a qubit is present can easily be
determined without affecting the qubit's state. The main result of
this paper is that the maximum tolerable erasure error rate can be as
high as $0.292$ per operation, given otherwise standard (though not
necessarily practical) scalability assumptions. Interestingly, the
only operations that have to meet this error probability are those
needed for storing a qubit for one time-step, and Bell measurements.
Other operations need have only a non-zero probability of success. If
qubit memory is perfect, any probability of erasure below $1/2$ during
a Bell measurement can be tolerated.

To establish a bound on threshold erasure error rates, the techniques
used in~\cite{knill:qc2000e} are adapted to the general setting.
Basically, all computational operations with error correction are
combined into a single, very flexible teleportation step.  The ideas
that make this possible can be found
in~\cite{steane:qc1999a,gottesman:qc1999a,steane:qc2002a,steane:qc2002b,steane:qc2003a}.
Which operation is applied and the means for error correction are
determined by which state is prepared for use in the teleportation
step. This is where error detection can be used to advantage: The
prepared state can be guaranteed to be error-free at the time it is
brought into the computation. Specifically, one can configure the
computation so that all states that may be needed are manufactured in
large quantities at a state factory, with any states for which errors
are detected being discarded before use.  Although this is extremely
inefficient, it means that for theoretical scalability, only errors in
the implementation of the teleportation process itself are relevant.

The techniques described here reduce the problem of fault-tolerant
quantum computation to the problem of preparing certain states with
sufficiently low error rates.  Assuming that storage and Bell
measurement errors are small enough, this analysis shows that
depolarizing error rates below $0.1$ per qubit are tolerable in the
prepared states. How this affects the maximum tolerable depolarizing
error at the level of individual quantum gates depends crucially on
how the states can be prepared.  For example, if there is a
sufficiently fault-tolerant way of preparing the states with about 100
gates contributing to the final error in each qubit, then one would
expect that a depolarizing error rate of $10^{-3}$ can be tolerated,
which is comparable to the thresholds shown
in~\cite{steane:qc2002a}. There is hope that fault-tolerant state
preparation can be accomplished with sufficiently low overhead, for
example by adapting and generalizing protocols for entanglement
purification~\cite{duer:qc2003a}.

\section{Error Models}

Threshold error rates depend in subtle ways on the details of the
error model adopted and how it is tied to the universal gate set used
for computation. For pure detected-error models, each gate or other
operation is error-free if no error is detected.  In the erasure error
model, a detected error implies complete loss of the state of the
qubits involved. This is the model focused on in this paper.  The error rate
is determined by the probabilities of detected error for the various
operations. For simplicity, all operations are assumed to take the
same amount of time (``one time-step'') and are synchronously clocked
on the qubits. It is assumed that classical computations are
instantaneous and error-free, and that operations can be applied to any
pair of qubits without communication delays.  In other words, there is
no communication or classical computation latency. Classical
computation latency can be a problem if the algorithms for calculating
the necessary error-correction steps depending on measurement outcomes
are too complex. Quantum communication latency occurs when distant
qubits need to be coupled. If classical computation latency is
negligible, quantum communication latency can be significantly reduced
by using teleportation methods~\cite{brennen:qc2003a}.

It is assumed that the probabilities of detected error are strictly
less than 1. The following detected-error probabilities will be used:
\begin{itemize}
\item[1.] $e_m$, the probability of detected error for the ``memory'' operation,
which involves storing the state of a qubit for one time-step.
\item[2.] $e_b$, the probability of detected error in implementing a Bell measurement
on two qubits.
\end{itemize}
Interestingly, probabilities of detected errors for other operations
do not affect the threshold if the methods described below are used.
It is assumed that errors are independent between different operations.

A reasonable model for unknown errors that are not detected is the
depolarizing error model. For a one-qubit operation, the state is
randomly affected by $\sigma_x$, $\sigma_y$ or $\sigma_z$.  The
probability that one of the Pauli operators ``occurs'' as an error is
the error probability.  For two-qubit operations the interpretation of
the model varies. One can assume that each of the possible
non-identity Pauli products occurs with equal probability,  but in this
work only storage and Bell-measurement errors will play a significant
role. For storage errors, let $d_m$ be the probability that one
of the non-identity Pauli matrices occurs. For Bell-measurement errors,
the effect is as if the correct measurement is physically applied, but
sometimes a random incorrect answer is learned. Let $d_b$ be the
probability of getting an incorrect answer in a Bell measurement.

\section{Stabilizer Codes}

Here is a brief review of the relevant stabilizer code theory.
Stabilizer codes for qubits are defined as a common eigenspace of a
set of commuting products of Pauli operators.  Let $n$ be the length
of the code, that is, the number of qubits used. It is convenient to
specify a product of Pauli operators (``Pauli products'') by a pair of
length-$n$ binary (row) vectors $\mathbf{s}=(\mathbf{a},\mathbf{b})$,
$\mathbf{a}=(a_i)_{i=1}^n, \mathbf{b}=(b_i)_{i=1}^n$.  For example,
consider $(a_1,a_2,a_3)=(1,0,1)$ and $(b_1,b_2,b_3)=(0,1,1)$.  For
brevity, one can omit commas and use square brackets as follows:
 $(a_1,a_2,a_3)=[a_1a_2a_3]=(1,0,1)=[101]$.  The product of
Pauli operators associated with $\mathbf{s}$ is given by
\begin{equation}
P(\mathbf{s})=P(\mathbf{a},\mathbf{b})=\prod_{j=1}^n
      \left\{\begin{array}{ll}
               \slb{\sigma_x}{j} &\textrm{if $a_j=1$ and $b_j=0$}\\
               \slb{\sigma_z}{j} &\textrm{if $a_j=0$ and $b_j=1$}\\
               \slb{\sigma_x}{j}\slb{\sigma_z}{j} = -i\slb{\sigma_y}{j}
                     &\textrm{if $a_j=1$ and $b_j=1$}
             \end{array}\right..
\end{equation}
\ignore{
   Note that for qudits with $d$ an odd prime, 
   the generators are $R$ (rotate) and $P$ (phase),
   and $P(\mathbf{s})= \prod_i R^{a_i}P^{b_i}$. For qubits,
   there is this extra phase of $i$ playing a role when $a_i=b_i=1$.

   One can try to simplify as follows: 
   Define $P$ in general by $P(\mathbf{s})= \prod_i R^{a_i}P^{b_i}$
   but introduce $\nu(\mathbf{s})$ as $e^{i\theta}$, where $\theta\geq 0$  is the
   the smallest angle such that $e^{i\theta}$ is an eigenvalue of $P(\mathbf{s})$.
   In the binary case, $\nu(\mathbf{s}) = i$ if $\mathbf{s}$ has an odd number
   of coordinates with $a_i=1$ and $b_i=1$. This function will be needed
   when defining binary syndromes and computing the associated eigenvalues.
}
Here, parenthesized superscripts denote the label of the qubit on which
the given operator acts.  Let $\mathbf{t}=(\mathbf{c},\mathbf{d})$
be another pair of length-$n$ binary vectors. All computations
involving vectors and matrices are performed modulo 2, with one
exception pointed out below.  $P(\mathbf{s})$ commutes with
$P(\mathbf{t})$ if $\mathbf{a}\mathbf{d}^T-\mathbf{b}\mathbf{c}^T=0$,
and anticommutes otherwise. (Although the minus sign in the identity
has no effect because arithmetic is modulo 2, it is retained for
consistency with the theory of non-binary stabilizer codes.)
Explicitly, $P(\mathbf{s})P(\mathbf{t}) =
(-1)^{\mathbf{a}\mathbf{d}^T-\mathbf{b}\mathbf{c}^T}P(\mathbf{t})P(\mathbf{s})$.
One can consider pairs of $n$-dimensional row vectors such as
$\mathbf{s}$ as $2n$-dimensional row vectors. To maintain the
association with qubit positions, it is convenient to merge the two
$n$-dimensional vectors.  That is, by definition,
$\mathbf{s}=(\mathbf{a},\mathbf{b})=[a_1b_1a_2b_2\ldots]$.  With the
example above, one can write $\mathbf{s}=[100111]$.  To make the
association with Pauli matrices, the abbreviations $I=00$, $X=10$,
$Z=01$ and $Y=11$ are convenient. Thus $[100111]=[XZY]$.  Note the
factor of $-i$ in the Pauli operator associated with $Y$.

\newcommand{\Sy}{\mathbb{S}}
Let $\Sy_l$ and $\Sy$ be the $2n\times 2n$ block-diagonal
matrices
\begin{equation}
\Sy_l = \left(
  \begin{array}{ccccc}
    0&0&0&0&\ldots\\
    1&0&0&0&\ldots\\
     0&0&0&0&\ldots\\
     0&0&1&0&\ldots\\
     \vdots&\vdots&\vdots&\vdots&\ddots
  \end{array}
\right), \;\;\;
\Sy=\Sy_l-\Sy_l^T=\left(
  \begin{array}{ccccc}
    0&-1&0&0&\ldots\\
    1&0&0&0&\ldots\\
     0&0&0&-1&\ldots\\
     0&0&1&0&\ldots\\
     \vdots&\vdots&\vdots&\vdots&\ddots
  \end{array}
\right).
\end{equation}
Then $P(\mathbf{s})$ commutes with
$P(\mathbf{t})$ if $\mathbf{s}\Sy\mathbf{t}^T=0$ and anticommutes
otherwise. Specifically, 
\begin{equation}
P(\mathbf{s})P(\mathbf{t}) = (-1)^{\mathbf{s}\Sy\mathbf{t}^T} P(\mathbf{t})P(\mathbf{s}).
\end{equation}
More generally, the multiplication rule is
\begin{equation}
P(\mathbf{s})P(\mathbf{t}) = (-1)^{\mathbf{s}\Sy_l\mathbf{t}^T}P(\mathbf{s}+\mathbf{t}).
\end{equation}

Consider $l$ independent $2n$-dimensional binary vectors
$\mathbf{s}_i$ describing Pauli products as explained above.  Let $Q$
be the matrix whose rows are the $\mathbf{s}_i$.  If the
$P(\mathbf{s}_i)$ commute, then the state space of $n$ qubits
decomposes into $2^l$ disjoint common eigenspaces of the
$P(\mathbf{s}_i)$, each of dimension $2^{n-l}$.  The eigenspaces are
characterized by their syndromes, that is, by the eigenvalues of the
$P(\mathbf{s}_i)$.  To standardize the eigenvalues in terms of binary
vectors requires introducing the function $\nu(\mathbf{s})=i^{\mathbf{s}\Sy_l\mathbf{s}^T}$, where expressions in the exponent of $i$ are not reduced
modulo $2$ (they can be reduced modulo $4$).
Note that $\mathbf{s}\Sy_l\mathbf{s}^T$ counts the number of
$Y$'s in $\mathbf{s}$, and each $Y$ contributes factors of $i$ to
the eigenvalues.
With this definition, $\nu([X])=\nu([Z])=1$, and $\nu([Y])= i$.
\ignore{
  Alternatively, one can associate $[Y]$ directly with $\sigma_y$, but that
  rather complicates the multiplication rules. This is really a problem with
  characteristic 2: For other characteristics, the function $\nu$ is trivial. 
  If $[Y]$ is defined as $\sigma_y$, then the multiplication
  rule is
  \begin{equation}
    P(\mathbf{s})P(\mathbf{t}) = (-1)^{\mathbf{s}\Sy_l\mathbf{s}^T+
                                       \mathbf{t}\Sy_l\mathbf{t}^T}
                                  i^{\mathbf{s}\Sy\mathbf{t}^T} 
          P(\mathbf{s}+\mathbf{t}).
  \end{equation}
  The syndrome change rule becomes
  \begin{eqnarray}
    P(\mathbf{s})\ket{\psi} &=&
       (-1)^{\mathbf{x}\mathbf{e}^T}
       \; (-1)^{\sum_i x_i \mathbf{s_i}\Sy_l\mathbf{s_i}^T}
       \; i^{\sum_{i<j} x_ix_j\mathbf{s_i}\Sy\mathbf{s_j}^T}
       \ket{\psi}
  \end{eqnarray}
}
Each syndrome is described by an $l$-dimensional
binary (row) vector $\mathbf{e}$  and relates
to the eigenvalues as follows: If $\ket{\psi}$ is in the eigenspace
with syndrome $\mathbf{e}=(e_i)_{i=1}^l$, then
$P(\mathbf{s}_i)\ket{\psi}= (-1)^{e_i}\nu(\mathbf{s}_i)\ket{\psi}$.
If $\mathbf{s}$ is in the row span of $Q$, that is,
$\mathbf{s}=\mathbf{x}Q$, then the eigenvalue of $P(\mathbf{s})$ for this
syndrome is expressible as
\begin{eqnarray}
P(\mathbf{s})\ket{\psi} &=& 
       (-1)^{\mathbf{x}\mathbf{e}^T}
             \; (-1)^{\sum_{i<j} x_ix_j\mathbf{s}_i\Sy_l\mathbf{s}_j^T}
       \prod_i \nu(\mathbf{s}_i)^{x_i}
       \ket{\psi}\nonumber\\
&=&    (-1)^{\mathbf{x}\mathbf{e}^T}
       \;   (-1)^{\sum_{i<j} x_ix_j\mathbf{s}_i\Sy_l\mathbf{s}_j^T} 
       \;i^{\sum_i x_i \mathbf{s}_i \Sy_l\mathbf{s}_i^T}
       \ket{\psi}\nonumber\\
&=&    (-1)^{\mathbf{x}\mathbf{e}^T}
        \;  (-1)^{\sum_{i<j} x_ix_j\mathbf{s}_i\Sy_l\mathbf{s}_j^T} 
       \;i^{\sum_i x_i \mathbf{s}_i \Sy_l\mathbf{s}_i^T}
       \;i^{-\mathbf{s}\Sy_l\mathbf{s}^T}
       \nu(\mathbf{s})
       \ket{\psi}\nonumber\\
&=&    (-1)^{\mathbf{x}\mathbf{e}^T}
       \;   (-1)^{\sum_{i<j} x_ix_j\mathbf{s}_i\Sy_l\mathbf{s}_j^T} 
       \;i^{\sum_i x_i x_i\mathbf{s}_i \Sy_l\mathbf{s}_i^T}
       \;i^{-\sum_{ij} x_i x_j \mathbf{s}_i\Sy_l\mathbf{s}_j^T}
       \nu(\mathbf{s})
       \ket{\psi}\nonumber\\
&=&    (-1)^{\mathbf{x}\mathbf{e}^T}
       \;   (-1)^{\sum_{i<j} x_ix_j\mathbf{s}_i\Sy_l\mathbf{s}_j^T} 
       \;i^{-\sum_{i\not=j} x_i x_j \mathbf{s}_i\Sy_l\mathbf{s}_j^T}
       \nu(\mathbf{s})
       \ket{\psi}\nonumber\\
&=&    (-1)^{\mathbf{x}\mathbf{e}^T}
       \; i^{\sum_{i<j} x_ix_j\mathbf{s}_i\Sy_l\mathbf{s}_j^T} 
       \;i^{-\sum_{i>j} x_i x_j \mathbf{s}_i\Sy_l\mathbf{s}_j^T}
       \nu(\mathbf{s})
       \ket{\psi}\nonumber\\
&=&    (-1)^{\mathbf{x}\mathbf{e}^T}
        \;i^{\sum_{i<j} x_ix_j\mathbf{s}_i\Sy_l\mathbf{s}_j^T} 
        \;i^{-\sum_{i>j} x_i x_j \mathbf{s}_i\Sy_l\mathbf{s}_j^T}
       \nu(\mathbf{s})
       \ket{\psi}\nonumber\\
&=&   (-1)^{\mathbf{x}\mathbf{e}^T}
        \;i^{\sum_{i<j} x_ix_j\mathbf{s}_i(\Sy_l-\Sy_l^T)\mathbf{s}_j^T} 
       \nu(\mathbf{s})
       \ket{\psi}\nonumber\\
&=&   (-1)^{\mathbf{x}\mathbf{e}^T}
        \;i^{\sum_{i<j} x_ix_j\mathbf{s}_i\Sy\mathbf{s}_j^T} 
       \nu(\mathbf{s})
       \ket{\psi}\nonumber\\
&=&   (-1)^{\mathbf{x}\mathbf{e}^T}
        \;i^{\mathbf{x} \,\scalebox{.8}{ut}(Q\Sy Q^T)\mathbf{x}^T}
       \nu(\mathbf{s})
       \ket{\psi},
\end{eqnarray}
where $\textrm{ut}(Y)$ is the strictly upper-triangular part of $Y$.
The projection
operator onto the eigenspace with syndrome $\mathbf{e}$ is given by
\begin{equation}
\Pi(Q,\mathbf{e}) = \prod_i{1\over 2}(\one + (-1)^{e_i}\bar\nu(\mathbf{s}_i)P(\mathbf{s}_i)).
\end{equation}
\ignore{
  For qudits: $\prod_i{1\over d}(\one + \omega^{-e_i}\bar\nu(\mathbf{s}_i)P(\mathbf{s}_i) 
                         + \omega^{-2e_i}(\bar\nu(\mathbf{s}_i)P(\mathbf{s}_i))^2
                         + \ldots )$,
  which uses the fact that the eigenvalue of 
  $\bar\nu(\mathbf{s}_i)P(\mathbf{s}_i)$ is $\omega^{e_i}$.
}
If it is necessary to emphasize the dependence of the syndrome on $Q$,
it will be referred to as the $Q$-syndrome.  The eigenspaces are the
stabilizer codes associated with $Q$.  For $\mathbf{s}$ in the row
span of $Q$, $P(\mathbf{s})$ stabilizes the states of these stabilizer
codes up to a phase.  Such Pauli products form the stabilizer of the
codes.  Write $\Pi(Q)=\Pi(Q,0)$ and consider this to be the
fundamental stabilizer code associated with $Q$. The word
``fundamental'' will be omitted whenever possible. Furthermore,
$\Pi(Q)$ is used to refer both to the projection operator and to the
code as a subspace: The intended meaning will be clear from the context.

It is important to understand the effects of Pauli products on states
in a stabilizer code. One can verify that
$P(\mathbf{s})\Pi(Q,\mathbf{e})P(\mathbf{s})^\dagger =
\Pi(Q,\mathbf{e'})$, where $\mathbf{e'} = \mathbf{e}+\mathbf{s}\Sy Q^T$.
This implies that if $\ket{\psi}$ has syndrome $\mathbf{e}$, so that
$\Pi(Q,\mathbf{e})\ket{\psi}=\ket{\psi}$, then
$P(\mathbf{s})\ket{\psi}$ has syndrome $\mathbf{e}+\mathbf{s}\Sy Q^T$.
To see this, compute $P(\mathbf{s})\ket{\psi}=
P(\mathbf{s})\Pi(Q,\mathbf{e})\ket{\psi} =
\Pi(Q,\mathbf{e}+\mathbf{s}\Sy Q^T)P(\mathbf{s})\ket{\psi}$.

Let $C=C(Q)$ be the row span of $Q$. $C$ is a classical binary
code. If $Q'$ has the same row span as $Q$, then $C(Q')=C(Q)$, and the
set of stabilizer codes associated with $Q'$ is the same as that
associated with $Q$.  For understanding the error-correcting
properties of stabilizer codes, one has to look at $C^\perp$, the set
of vectors $\mathbf{x}$ such that $\mathbf{x}\Sy Q^T=0$, or,
equivalently, such that $P(\mathbf{x})$ commutes with all of the
$P(\mathbf{s_i})$. For $\mathbf{t}\in C^\perp$ but not in $C$,
$P(\mathbf{t})$ preserves each stabilizer code associated with $Q$ but
acts nontrivially in each code. Consequently, the quantum minimum
distance of these codes is the minimum distance of the set
$C^\perp\setminus C$. Here, minimum distance is defined as
the weight of the smallest-weight (non-zero) vector in $C^\perp\setminus C$.
The weight of $\mathbf{x}$ is the number of qubits on which
$P(\mathbf{x})$ acts nontrivially.

When working with stabilizer codes and syndrome measurements, it is
helpful to be able to determine the new stabilizer of a state after
making a syndrome measurement for a different code. Let $Q$ be as
above.  Suppose that the initial state $\ket{\psi}$ is an arbitrary
state of $\Pi(Q,\mathbf{e})$ and that one measures the $R$-syndrome
with outcome $\mathbf{f}$.  What Pauli products are
guaranteed to stabilize the resulting state $\ket{\phi}$?
$P(\mathbf{r})$ stabilizes $\ket{\phi}$ if $\mathbf{r}$ is in $C(R)$
or in $C(Q)\cap C(R)^\perp$. The latter set consists of the Pauli
operators guaranteed to stabilize the initial state that commute with
the measurement. In general, the only Pauli products guaranteed to
stabilize $\ket{\phi}$ are products of
the above. One can construct an independent set of such products from
$Q$ and $R$ by the usual linear-algebra methods modulo 2.  The
eigenvalues can be determined using the formulas introduced earlier.

It is not the case that minimum distance completely determines whether
$\Pi(Q)$ is a stabilizer code with good error-correction properties
for typical independent error models. That is, provided that the
number of low-weight elements of $C^\perp\setminus C$ is sufficiently
small, it is still possible to correct most errors. Suppose that
$\ket{\psi}$ is encoded as $\kets{\psi}{L}$ in $\Pi(Q)$. For any error
model, the effect of the errors on $\kets{\psi}{L}$ can be thought of
as a probabilistic mixture of the states $A_k\kets{\psi}{L}$, where
$(A_k)$ are the operators in the operator sum representation of the
errors and satisfy $\sum_k A_k^\dagger A_k=\one$. The probability of
$A_k\kets{\psi}{L}$ is $\bras{\psi}{L}A_k^\dagger A_k\kets{\psi}{L}$.
Because Pauli products form a complete operator basis,
$A_k=\sum_{\mathbf{s}}\alpha_{k\mathbf{s}}P(\mathbf{s})$.  To correct
the errors one can measure the $Q$-syndrome of the noisy
state. Suppose that the measured syndrome is $\mathbf{e}$.  Then the
state $A_k\kets{\psi}{L}$ is projected to
$\sum_{\mathbf{s}:\mathbf{s}\Sy Q^T =
\mathbf{e}}\alpha_{k\mathbf{s}}P(\mathbf{s})\kets{\psi}{L}$. The sum
is over a set $C^\perp +\mathbf{s}_0$. A good code for
the error model has the property that, with high probability, all
dominant amplitudes among the $\alpha_{k\mathbf{s}}$ satisfy 
the condition that $\mathbf{s}$
is in the same set $C+\mathbf{s}'\subseteq
C^\perp+\mathbf{s}_0$, independent of which $A_k$
occurred. If that is true, then a decoding algorithm can determine
the dominant amplitude's coset $C+\mathbf{s}'$ and apply
$P(\mathbf{s}')^\dagger$ to restore $\kets{\psi}{L}$. A practical code
also has the property that there is an efficient decoding algorithm
that has a high probability of successfully inferring $C+\mathbf{s}'$.

The discussion of the previous paragraph assumes that nothing is known
about the error locations. Suppose that it is known that the errors
occurred on a given set $S$ of $m$ qubits.  If the errors are
erasures, without loss of generality, reset the erased qubit to $0$
(replacing it with a fresh qubit if necessary).  Suppose that after
this, the measured syndrome is $\mathbf{e}$.  The possible Pauli
products appearing in the new state
$\ket{\phi}=\sum_{\mathbf{s}:\mathbf{s}\Sy Q^T =
\mathbf{e}}\alpha_{k\mathbf{s}}P(\mathbf{s})\kets{\psi}{L}$ satisfy
the condition that $\mathbf{s}$ has non-zero entries only for qubits
in $S$ and $\mathbf{s}\in C^\perp +\mathbf{s_0}$ for some
$\mathbf{s_0}$.  Suppose that $C^\perp\setminus C$ contains no
$\mathbf{s}$ with non-zero entries only for qubits in $S$.  Then all
$\mathbf{s}$ appearing in the sum for $\ket{\phi}$ are in the same set
$C+\mathbf{s}'$ for some $\mathbf{s}'$.  Applying $P(\mathbf{s}')$
corrects the error. Note that a suitable $\mathbf{s}'$ can be computed
efficiently given $\mathbf{e}$ and $S$: It suffices to solve
$\mathbf{s}\Sy Q^T = e$ subject to the condition that $\mathbf{s}$ is
zero for positions associated with qubits outside of $S$.  This is a
set of linear equations modulo $2$.  

Because of the argument of the previous paragraph, an erasure code for
$S$ is defined as a code $C$ such that $C^\perp\setminus C$ contains
no $\mathbf{s}$ with nonzero entries only for qubits in $S$. It
follows that a code of minimum distance $d$ is an erasure code for all
$S$ of cardinality at most $d-1$. A useful property of erasure codes
when all errors are detected is that if an error combination cannot be
corrected, then this is known. This is because given $S$ it is
possible to determine whether the code is an erasure code for $S$. If
an error combination cannot be corrected, this becomes a detected
error for the encoded information. In particular, for the erasure
error model, the encoded information is also subject to erasure errors
(hopefully at a much lower rate). In other words, the error model is
preserved by encoding.

In addition to being able to correct errors with high probability, a
good stabilizer code should be able to encode a large number of
qubits. For the present purposes, analysis is simplified by encoding
one qubit at a time. However, efficiency can be improved substantially
by encoding more and the basic techniques that are used are still
applicable.  Let $Q$ be a matrix with $n-1$ rows defining a
two-dimensional stabilizer code. A qubit can be encoded in a way
consistent with the stabilizer formalism by choosing two row vectors
$\mathbf{t}_x$ and $\mathbf{t}_z$ with the property that
$\mathbf{t}_x\Sy Q^T =\mathbf{0}$, $\mathbf{t}_z\Sy Q^T=\mathbf{0}$
and $\mathbf{t}_x\Sy\mathbf{t}_z^T = 1$. Then $P(\mathbf{t}_x)$ and
$P(\mathbf{t}_z)$ relate to each other as $X$ and $Z$ and can
therefore serve as encoded $X$ and $Z$ observables. Note that if $Q$
is extended by $\mathbf{t}_x$, $\mathbf{t}_z$ or
$\mathbf{t}_y=\mathbf{t}_x+\mathbf{t}_z$, then one-dimensional
stabilizer codes are obtained whose states are encoded $X$, $Z$ and
$Y$ eigenstates.

\section{Error Correction by Teleportation}

Let $Q$ be the $l\times 2n$ matrix defining a stabilizer code for encoding
$k=n-l$ qubits with good error-correction properties.  Consider $n$
qubits carrying a state encoded in the stabilizer code for $Q$ that
has been affected by errors. An effective way of correcting errors in
this state is to teleport each of the qubits using $n$ pairs of qubits
prepared as follows: First place each pair in the standard Bell state
$(\ket{00}+\ket{11})/\sqrt{2}$. Use a $Q$-syndrome measurement on the
$n$ second members of each pair to project them into one of the
stabilizer codes associated with $Q$. Finally, apply identical Pauli
matrices to both members of pairs in such a way as to reset the
syndrome to $0$. The result is that the $n$ pairs are in a state where
the first and second members are in a maximally entangled state of the
stabilizer code for $Q$. Teleportation in the absence of errors 
transfers the state of the input qubits to the output qubits.  In the
presence of errors, the Bell measurements used for teleportation
reveal syndrome information that can be used to correct some
errors. The remainder of this section is dedicated to establishing the
details of this procedure.

The standard quantum teleportation protocol begins with an arbitrary
state $\kets{\psi}{1}$ in qubit $\sysfnt{1}$ and the Bell state
$(\kets{00}{23}+\kets{11}{23})/\sqrt{2}$ on qubits
$\sysfnt{2},\sysfnt{3}$. The initial state can be viewed as
$\ket{\psi}$ encoded in the stabilizer code generated by
$\mathbf{b}_1=[001010]=[IXX]$ and $\mathbf{b}_2=[000101]=[IZZ]$.  
\ignore{
For odd characteristic, use $[001010]$ and $[00010(-1)]$.
}
Let $\slb{B}{23}$ be the matrix whose rows are the $\mathbf{b}_i$.
The stabilizer consists of the Pauli products
$\one,\slb{\sigma_x}{2}\slb{\sigma_x}{3},\slb{\sigma_y}{2}\slb{\sigma_y}{3}$
and $\slb{\sigma_z}{2}\slb{\sigma_z}{3}$.  To teleport, one makes a
Bell-basis measurement on the first two qubits.  This is equivalent to
making a $\slb{B}{12}$-syndrome measurement, where $\slb{B}{12}$ has
as rows $[101000]=[XXI]$ and $[010100]=[ZZI]$.  This is identical to
$\slb{B}{23}$ with qubits $\sysfnt{2},\sysfnt{3}$ exchanged for qubits
$\sysfnt{1},\sysfnt{2}$. Depending on the syndrome $\mathbf{e}$ that
results from the measurement, one applies correcting Pauli matrices to
qubit $\sysfnt{3}$ to restore $\ket{\psi}$ in qubit $\sysfnt{3}$.

One way to determine the required corrections is to follow the
procedure for $\kets{\psi}{1}$ being eigenstates of $\sigma_x$ and
$\sigma_z$ and to check which Pauli operators stabilize the final
state and what their eigenvalues are. Formally, start with a
one-dimensional stabilizer code generated by $B([ab])$ whose rows are
$[001010], [000101], [ab0000]$. Define $\mathbf{x}=[ab]$.  The initial
syndrome and $\mathbf{x}$ determine the input state on the first
qubit.  Suppose that the initial state has $B(\mathbf{x})$-syndrome
$\mathbf{e}=[00e]$.  Measure the $\slb{B}{12}$-syndrome,
obtaining syndrome $\mathbf{f}=[f_1f_2]$.  Among the stabilizers of
the new state is $P(\mathbf{s})$ with
$\mathbf{s}=a[001010]+b[000101]+[ab0000]+a[101000]+b[010100]=[0000ab]$.
\ignore{For odd characteristic:
  $a[001010]-b[00010(-1)]+[ab0000]-a[101000]-b[010(-1)00]$
}
The first three terms are in the row span of $B(\mathbf{x})$ and yield
a Pauli product commuting with the measurement.  The last two terms
are in the row span of $\slb{B}{12}$.  Thus the new state has the same
stabilizer on qubit $\sysfnt{3}$ as the original state had on qubit
$\sysfnt{1}$. However, the eigenvalue may be different. It
is given by $(-1)^{e+\mathbf{f}[ab]^T}i^{ab}$ (the exponent of $i$ is
the product of $a$ and $b$.) 
\ignore{For odd characteristic:
$(-1)^{e-\mathbf{f}[ab]^T}$.}
To restore the original eigenvalue,
it suffices to apply $P(\mathbf{f}\Sy)$ to qubit $\sysfnt{3}$.
\ignore{For odd characteristic:
$P(-\mathbf{f}\Sy)$}

The protocol for error correction by teleportation has $n$ qubits in a
state that initially was stabilized by $Q$ with syndrome
$\mathbf{0}$. The goal is to measure the $Q$-syndrome of the noisy
state $\ket{\psi}$ and obtain a corrected state at the teleportation
destination.  Label the $n$ qubits by $\sysfnt{1},\ldots,\sysfnt{n}$.
Adjoin $2n$ qubits labeled $\sysfnt{n+1},\ldots,\sysfnt{3n}$.  These
qubits are prepared as follows: Initialize qubits $\sysfnt{n+k}$ and
$\sysfnt{2n+k}$ in the Bell state (stabilizer
$\slb{B}{(n+k)(2n+k)}$). Measure the $Q$-syndrome on qubits
$\sysfnt{2n+1},\ldots,\sysfnt{3n}$. Make this syndrome identically $0$
by applying a Pauli product to these qubits after the
measurement. Apply the same Pauli product to the corresponding qubits
$\sysfnt{n+1},\ldots,\sysfnt{2n}$.  The state of qubits
$\sysfnt{n+1},\ldots,\sysfnt{3n}$ is stabilized by
$[\mathbf{0}\mathbf{s}_i]$ for the rows $\mathbf{s}_i$ of $Q$ and by
$[\mathbf{r}\mathbf{r}]$ for $\sysfnt{r}\in C(Q)^\perp$.  (Here, the
square bracket notation has been adapted to denote concatenation of
row vectors.) Note that because $C(Q)\subseteq C(Q)^\perp$, the
stabilizer also contains $[\mathbf{s}_i\mathbf{0}]$, so an equivalent
state could have been prepared by measuring the qubits
$\sysfnt{n+1},\ldots,\sysfnt{2n}$. Finally, teleport qubits
$\sysfnt{m}$ to $\sysfnt{2n+m}$ using the Bell measurement of qubits
$\sysfnt{m}$ and $\sysfnt{n+m}$. The claim is that the Bell
measurement results $\mathbf{f}_{m}$ are constrained in a way that
determines the $Q$-syndrome of the input state. The output state on
qubits $\sysfnt{2n+1},\ldots \sysfnt{3n}$ is the result obtained after
measuring the $Q$-syndrome of the input state on qubits
$\sysfnt{1},\ldots,\sysfnt{n}$.  Let $P(\mathbf{g})$ be the
Pauli-product correction applied to qubits
$\sysfnt{2n+1},\ldots,\sysfnt{3n}$ as part of the protocol.  The proof
of the claim is presented pictorially in Fig.~\ref{fig:teleec}.  It
shows that the $Q$-syndrome $\mathbf{e}$ is determined by the
measurement outcomes according to $\mathbf{e}=\mathbf{g}\Sy Q^T$.

\begin{herefig}
\label{fig:teleec}
\begin{picture}(0,7.2)(0,-6.9)
\nputgr{-3,-1}{l}{height=2.4in}{teleec_a}
\nputbox{-3.3,.1}{lt}{$n$ input qubits}
\nputbox{0.1,-.97}{lt}{$n$ output qubits}
\nputbox{-2.3,-1.04}{c}{\rotatebox{-90}{Bell}}
\nputbox{-1.58,-1.31}{c}{\small\rotatebox{-90}{$\Pi(Q,0)$}}
\nputbox{-.82,-.5}{c}{\rotatebox{90}{Bell}}
\nputbox{-.21,-1.27}{c}{$P(\mathbf{g})$}

\nputbox{-1.7,-3.3}{r}{\scalebox{2.5}{$\Leftrightarrow$}}
\nputgr{-1.5,-3.3}{l}{height=2.4in}{teleec_b}
\nputbox{-.8,-3.3}{c}{\rotatebox{-90}{Bell}}
\nputbox{-.05,-2.75}{c}{\rotatebox{90}{Bell}}
\nputbox{.55,-3.56}{c}{$P(\mathbf{g})$}
\nputbox{1.25,-3.56}{c}{\small\rotatebox{-90}{$\Pi(Q,\mathbf{g}\Sy Q^T)$}}

\nputbox{-.2,-5.6}{r}{\scalebox{2.5}{$\Leftrightarrow$}}
\nputgr{0,-5.6}{l}{height=2.4in}{teleec_c}
\nputbox{2.75,-5.88}{c}{\small\rotatebox{-90}{$\Pi(Q,\mathbf{g}\Sy Q^T)$}}
\end{picture}
\nopagebreak
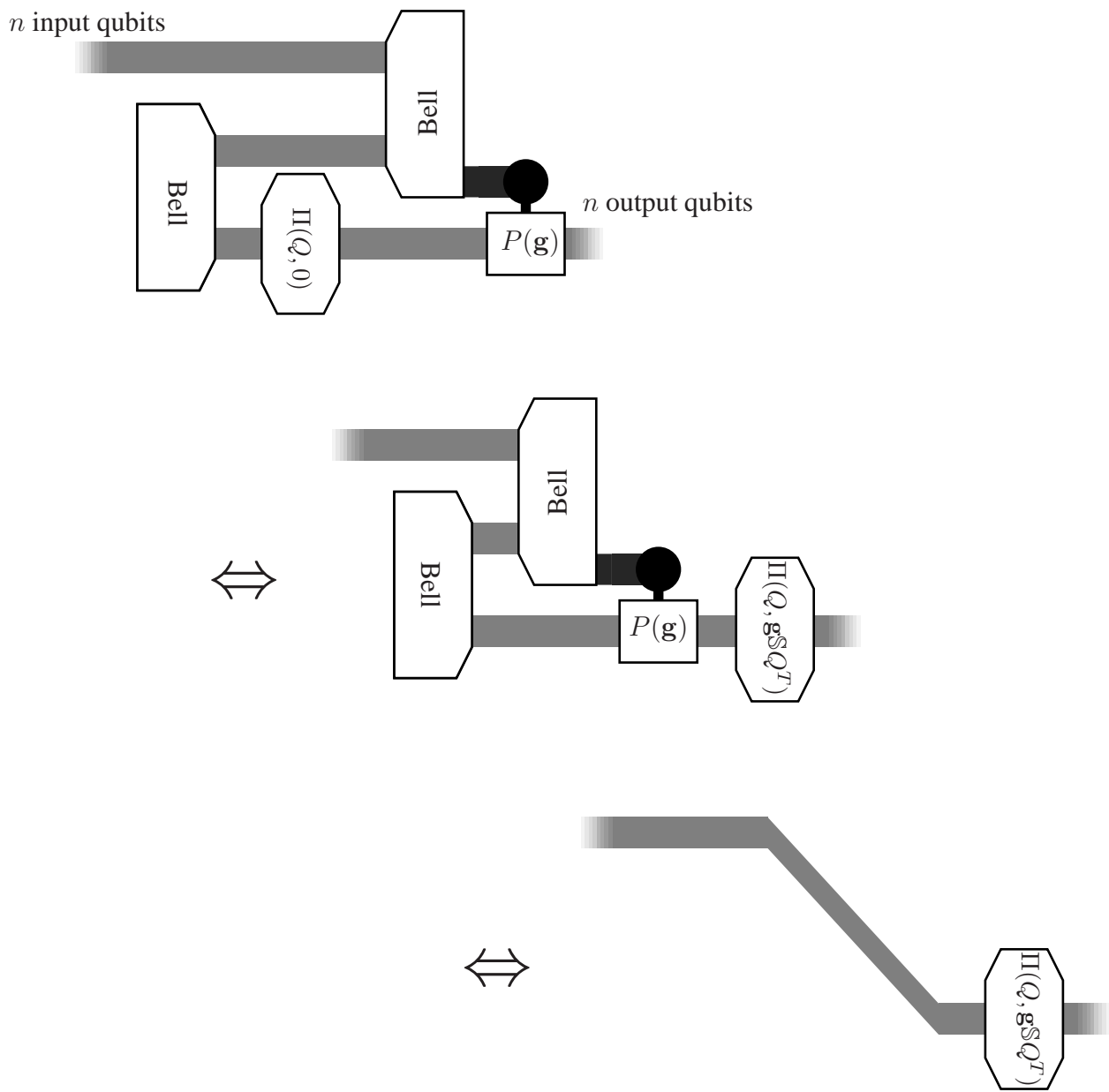
\herefigcap{Teleporting with an encoded entangled state is equivalent
to a syndrome measurement. The principle is as explained
in~\cite{gottesman:qc1999a}. In this case, a stabilizer projection on
the destination qubits before teleportation is equivalent to a
projection after teleportation, where the syndrome associated with the
projection is modified by the correction Pauli product used during
teleportation. The equivalence is shown with a transformation of
quantum networks in three steps. The gray lines are the time lines of
$n$ qubits. The boxes denote various operations.  The Bell-state
preparation on corresponding pairs of qubits in two sets of $n$ qubits
is depicted with a box angled to the right and labeled ``Bell''.  The
prepared state is the state obtained by projecting the
destination qubits with $\Pi(Q,0)$. Projection operators are shown
with boxes angled both ways with the operator written in the box. To
bypass the problem that a projection operator cannot be applied with
certainty, the Bell-state measurement followed by the projection may
be implemented by a suitable stabilizer code encoding procedure or the
procedure described in the text.  The Bell-state measurement is
depicted with a box angled to the left.  The measurement outcome is
carried by the darker, classical line exiting at the bottom of the
box. The Pauli-product correction operator is controlled by the
classical line. The conclusion is that the $Q$-syndrome can be
determined from the Pauli-product correction $P(\mathbf{g})$ by
computing $\mathbf{g}\Sy Q^T$.}
\end{herefig}

It does not matter which projection $\Pi(Q,\mathbf{t})$ is
applied after the Bell-state preparation. Provided $\mathbf{t}$ is
known, this affects only the deduced $Q$-syndrome for the input state,
which is now computed as $\mathbf{g}\Sy Q^T+\mathbf{t}$. As a result,
one can prepare the state needed for teleportation by measuring
the $Q$-syndrome of the destination qubits before teleporting,
and recording the outcome without compensating for it.

To use the above procedure for error correction, it suffices to modify
the Pauli products needed for completing the teleportation step by
multiplying with the Pauli products that are needed to compensate for error
according to the syndrome. Suppose that the quantum information of
interest was encoded with syndrome $\mathbf{0}$ before errors.
Suppose that the true error is (close to) $P(\mathbf{d}_t)$ If the
Bell measurements are error-free, then the syndrome after errors is
$\mathbf{g}\Sy Q^T$ and the optimal correction can be determined from
this, usually by finding a low weight $\mathbf{h}$ such that
$\mathbf{h}\Sy Q^T=\mathbf{g}\Sy Q^T$.  For good codes,
$\mathbf{h}=\mathbf{d}_t$ with high probability.  Suppose that there
are errors in the Bell measurement and instead of the error-free
$\mathbf{g}$, $\mathbf{g}'$ is used for correction.  Assuming
independence, the errors are still local, and one expects
$\mathbf{d}=\mathbf{g}'-\mathbf{g}$ to have low weight.  Because the
teleportation correction is incorrect, an additional error
$P(\mathbf{d})$ has been introduced.  That is, the new error is
$P(\mathbf{d}+\mathbf{d}_t)$. Fortunately, the computed syndrome
$\mathbf{g}'\Sy Q^T$ reflects this error, and the inferred correction
will work provided that the code is good for the combination of the
pre-Bell-measurement error process and the errors in the Bell
measurement.

It appears that erasure errors cause a problem because for cases where
such errors occurred the Bell measurement outcomes are unknown. The
argument of the previous paragraph shows that the way in which the
teleportation correction is ``filled in'' does not affect the outcome
of the procedure.  The correct state is reproduced at the
teleportation output in each case.

It is worth making a few simplifications to reduce the gate
overhead. One is to omit the Pauli products needed for teleportation
and error correction by using classical bookkeeping to keep track of
the current syndrome and the current ``Pauli frame'' for the encoded
information. One can reduce the bookkeeping problem by deferring the
Pauli correction (for both teleportation and errors) to the state
prepared for the next teleportation operation.

\section{Combining Operations with Error Correction}

Operations can now be integrated into the error-correction process
using the techniques described in~\cite{gottesman:qc1999a}.  The basic
idea is to apply the desired encoded operation to the destination
qubits of the entangled state (or states) to be used for
teleportation.  If the operation is in the so-called Clifford group,
the teleportation protocol results in the desired operation being
applied, except that the Pauli products needed to correct the state
are modified. To achieve universality, an additional operation that
has the property of conjugating Pauli products to elements of the
Clifford group is required.  One such operation is the $45^\circ$
rotation $e^{-i\sigma_x\pi/8}$. Again, it is applied in encoded form
to the destination qubits of the prepared state used for
teleportation. After teleportation, the necessary correction may be a
Clifford-group element not of the form of a Pauli product.  If that is
the case, this Clifford-group element is applied in the next
teleportation step. 

Note that for applying Clifford-group elements such as the
controlled-not, the teleportation step has to act on two encoded
qubits and the error-correction aspects of the teleportation step for
the two qubits have to be implemented on both at the same time.

\section{Measurement}

The same teleportation step used for computation can also be used for
measurement, except that in this case the destination qubits are
redundant. That is, if one prepares an encoded state on the source
qubits with the encoded qubits in logical zero, the Bell measurement
will reveal not just the syndrome of the code, but also the
measurement outcome. Errors can be classically corrected to reveal the
true measurement outcome.

\section{Thresholds}

The sequential implementation of the scheme in the context of a
computation is shown in Fig.~\ref{fig:full_scheme}. 

\begin{herefig}
\label{fig:full_scheme}
\begin{picture}(0,5)(0,-4.7)
\nputgr{-.25,0}{t}{height=4.4in}{teleseq}
\nputbox{-1.35,-1.25}{c}{\rotatebox{-90}{\begin{tabular}{c}State\\Preparation\end{tabular}}}
\nputbox{-.5,-.68}{c}{\rotatebox{90}{\begin{tabular}{c}Transversal Bell\\Measurement\end{tabular}}}

\nputbox{-.2,-2.4}{c}{\rotatebox{-90}{\begin{tabular}{c}State\\Preparation\end{tabular}}}
\nputbox{.68,-1.83}{c}{\rotatebox{90}{\begin{tabular}{c}Transversal Bell\\Measurement\end{tabular}}}

\nputbox{1,-3.55}{c}{\rotatebox{-90}{\begin{tabular}{c}State\\Preparation\end{tabular}}}
\end{picture}
\nopagebreak
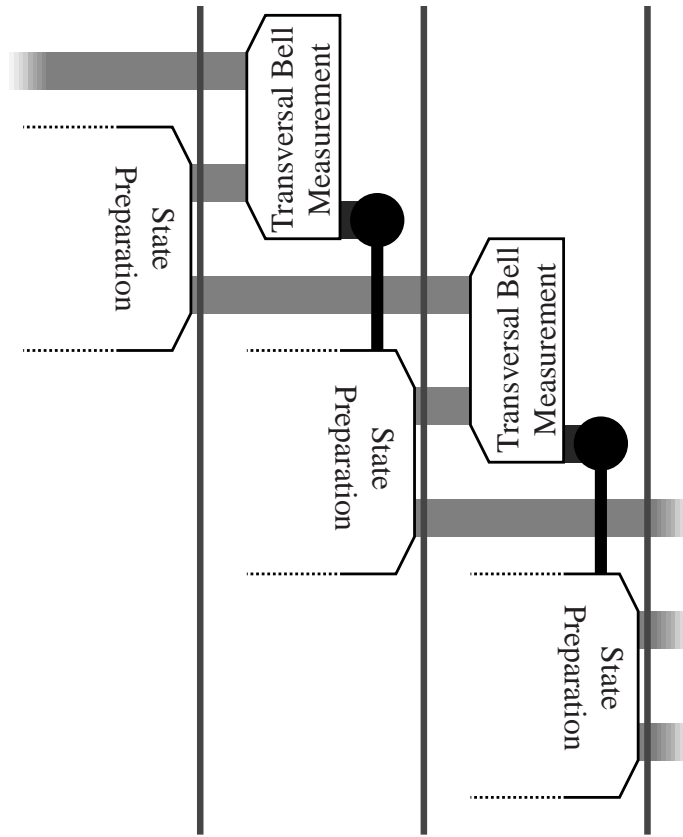
\herefigcap{Sequence of computational steps. Each step consists of a
teleportation with integrated operations and error correction.  The
steps required for correction determine the prepared state used in the
next step. That is, a state that has the appropriate operations
pre-applied to the destination qubit is used. Any such state is
assumed to be available at the ``state preparation factory'' at the
next step.  Zero communication and classical computation delays are
assumed.  The prepared states have been checked for errors, with no
errors detected just before it is used. Errors in the computation are
therefore due only to the Bell measurement and the storage time for
the destination qubits used during the Bell measurement.}
\end{herefig}

In order to obtain a lower bound on the erasure error threshold and
discussing the depolarizing error threshold, observe that the
efficiency with which the needed states are prepared has no effect on
the threshold. It is necessary only to determine for what error rates
it is possible to make the error per step in the encoded (logical)
qubits arbitrarily small.  Consider the erasure error model. In this
case, the error model for the logical qubits is also independent
erasures. If the rate of erasures is sufficiently small, then
according to the known threshold theorems, we can use the logical
qubits to efficiently implement arbitrarily accurate quantum
computations. Efficiency in these theorems requires only that the cost
of each elementary step of the computation is bounded by a constant
independent of the length of the computation.  Here, this constant
depends on the length of the code needed to achieve an encoded error
below the general threshold, for which there are known lower bounds.
Here, the effort required to prepare the states for the teleportation
steps only adds to the constant. In particular, the states can be
prepared naively, by attempting to implement a quantum network that
prepares them, and discarding any unsuccessful attempts.  It is known
that any one-qubit state encoded in a stabilizer code of length $n$
requires at most $O(n^2)$ quantum gates~\cite{cleve:qc1996b}.  Because
each succeeds only with probability $s<1$, the expected number of
attempts for each state to be prepared is bounded by
$e^{O(n^2)}$. Although this is superexponential, it contributes
``only'' a constant overhead to the implementation of each encoded
operation. A similar argument can be made for the depolarizing error
model, except that the final analysis may be affected by the residual
(undetected) errors in the prepared states. In conclusion, it is
possible to invoke the general threshold theorems to show that state
preparation overhead can be ignored for the purpose of establishing
lower bounds on the threshold by the methods used here. Nevertheless,
a self-contained proof not relying on the general threshold theorems
of scalability is given in Sect.~\ref{sect:ineff}.

Suppose that we use a one-qubit erasure code for which the probability
of an uncorrectable erasure is $f(e)$, given that the probability of
erasure of each qubit is $e$.  The error probability of a quantum
computation using the scheme of Fig.~\ref{fig:full_scheme} is
determined by the probability that the Pauli product correction to
restore the state in the destination qubits is incorrectly
inferred. By the time this correction is inferred, the destination
qubits that now carry the desired state are of course already
corrupted. However, any errors that have occurred are removed in the
next step and therefore ignored in this analysis. The probability of
detected error is given by $e_m+(1-e_m)e_b$. The first term comes from
detected error in the storage period of the destination qubits. The
second comes from the Bell measurement, which is applied only to
qubits with no previously detected error. The probability of erasure
of the encoded qubit is therefore $f(e_m+(1-e_m)e_b)$. Because the
prepared states are error-free at the instant when they are used, this
erasure probability also applies to each qubit independently in
teleported two-qubit operations.  It remains to determine for what
error rates there exist one-qubit erasure stabilizer codes with
arbitrarily small probability of encoded erasure. If the supremum of
these error rates is $e_{\textrm{\tiny max}}$, then the threshold for
$e_m$ and $e_b$ is determined by the curve $e_{\textrm{\tiny max}} =
e_m+(1-e_m)e_b$. The value of $e_{\textrm{\tiny max}}$ is determined
in~\cite{bennett:qc1997b} and is given by $e_{\textrm{\tiny max}} =
1/2$. Fig.~\ref{fig:threshold} shows the region for $e_m$ and $e_b$
where scalable computation for the detected-error model is possible.

\ignore{
x=[.0:.01:.5];
y=max([min([(.5-x)./(1-x);ones(1,length(x))]);zeros(1,length(x))]);
gset term postscript
gset output "graphics/thcurve.ps"
plot(x,y);
}

\begin{herefig}
\label{fig:threshold}
\begin{picture}(0,4)(0,-3.7)
\nputgr{0,0}{t}{height=3.5in}{thregion}
\nputbox{2.28,-3.5}{tl}{$e_m$}
\nputbox{-2.35,-.08}{tr}{$e_b$}
\nputbox{-.8,-.3}{tl}{$e_m+(1-e_m)e_b=1/2$}
\nputbox{.28,-1.3}{bl}{$e_m=e_b=1-1/\sqrt{2}>0.292$}
\end{picture}
\nopagebreak
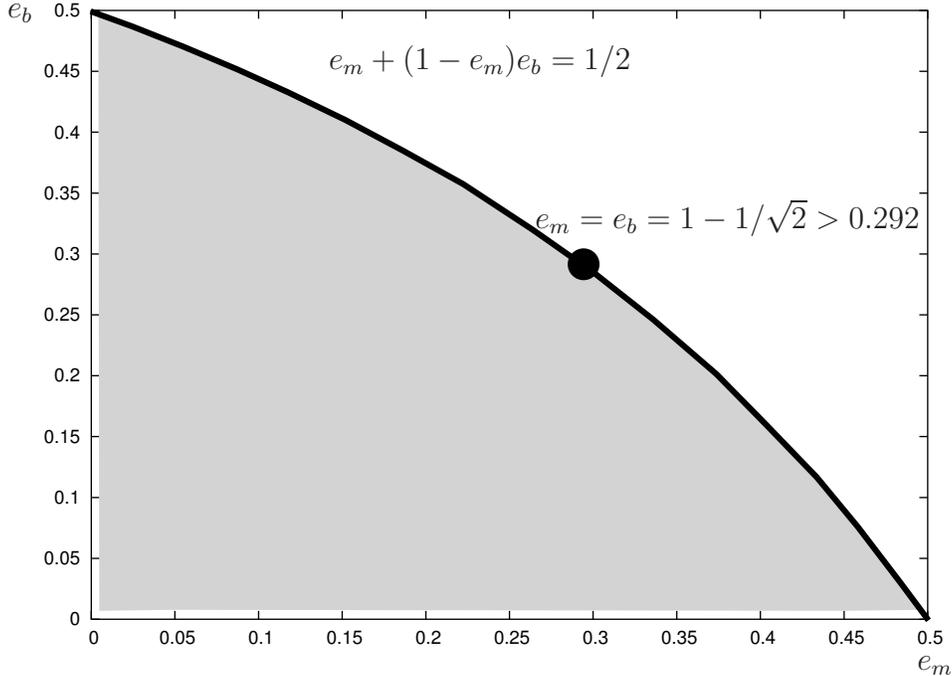
\herefigcap{Region for which scalable computation for the
detected-error model is possible. Tolerable error rates are in the
gray region, strictly below the upper boundary. If memory and Bell
measurement error are equal, then any error-rate below $0.292$ is
tolerable. The point on the boundary corresponding to this value is
shown. If memory errors are negligible, then Bell-measurement error
rates close to $1/2$ are tolerable.}
\end{herefig}

\section{State-Preparation Inefficiency}
\label{sect:ineff}

It is possible to prove that state preparation can be done with
polynomial overhead directly instead of relying on the general
threshold theorems. This is done here to make the
paper more self contained and to show that two levels of concatenation
suffice for the erasure error model. Let $n$ be the final length of
the (concatenated) code for each logical qubit. The code is
constructed by concatenating two erasure codes of length $l_1$ and
$l_2$ with $l_1l_2=n$ and $l_1<\sqrt{n}$. State preparation
through the first level of encoding is handled by the naive method of
repeated attempts. At the second level of encoding, the methods of the
previous sections are used to improve the probability of successful
state preparation. 

To see that one level of encoding with the naive state-preparation
method is insufficient, consider the following: The goal is to
implement a computation of $N$ elementary operations with polynomial
overhead. In order for the computation to have a probability
$1-\epsilon$ of success with logical qubits and no further error
correction, the logical qubits must be subject to an error rate of
$\epsilon/N$ per operation.  (Erasure errors add
probabilistically.) Given that $e_m$ and $e_b$ are in the scalability
region, the error rate as a function of code length $n$ for the best
codes goes as $e^{-cn}$ (asymptotically) for some constant $c$
depending on $e_m$ and $e_b$. One should therefore choose
$n>\ln(N/\epsilon)/c$.  Typically, the state-preparation networks for
these codes have at least $c'n^2$ gates for some constant $c'>0$. The
naive state preparation therefore requires resources of order
$e^{c''n^2}$. Substituting the lower bound for $n$ gives a
superpolynomial function of $N/\epsilon$.

To eliminate the superpolynomial overhead, choose a first level code
that reduces the error rate to $e^{-cl_1}$. Choose a second level code
that can correct any combination of at most $l_2/6$ erasures (such
stabilizer codes exist by using random
coding~\cite{bennett:qc1996a,calderbank:qc1996a}). The concatenation
of the two codes results in logical qubits with an error rate bounded
by ${l_2\choose l_2/6}(e^{-cl_1})^{l_2/6}\leq e^{-c'n}$ for some
constant $c'>0$ and sufficiently large $l_1, l_2$.  Choose
$n>\ln(N/\epsilon)/c'$.  The naive state-preparation method for
computations with qubits encoded in the first-level code requires an
overhead of $e^{c''l_1^2}\leq e^{dn}$ for some constant $d$.
Computations at the next level are implemented using the teleportation
techniques discussed above. The probability of success of one step is
at least $1-e^{-c'''l_1}$ for some constant $c'''$ (different from $c'$
because a step involves both storage and Bell measurements). The
probability of success of a state-preparation network for states
needed by the second level code is at least $1-c'l_2^2
e^{-c'''l_1}>1/2$ for sufficiently large $n$.  The total overhead for
the concatenated state preparation is bounded by $2c'l_2^2 e^{dn}<
e^{d'\ln(N/\epsilon)} = \mathrm{poly}(N/\epsilon)$.

An obvious way to improve the efficiency of this state-preparation
scheme is to choose the size of the first level code to better balance
the overheads between the two levels. In particular, the first level
code need have length only of order $\log(n)$ for the probability of
success of the second-level state preparation to exceed a
constant. This reduces the overhead to polylogarithmic in $N/\epsilon$,
comparable to overheads in the standard threshold theorems based on
concatenation.

\section{Application to other error models}

The techniques discussed above can be used with any error model
provided that it is possible to prepare the requisite state such that
the final error in the state is sufficiently well controlled.  For
example, consider the depolarizing error model.  Here, there are
probabilities for complete, undetected depolarization for the various
operations.  The probability of a depolarizing error is the
probability that some non-identity Pauli operator affects a qubit.
Suppose that the probability of a depolarizing error for a memory
operation is $d_m$ and that the probability of a random incorrect Bell
measurement outcome is $d_b$. Suppose also that encoded states can be
prepared such that they are as intended except for independent
depolarizing errors on each qubit with probability of error
$d_p$. Assume also that this probability does not depend on the code
used in a family of codes of interest. Such a family of codes could be
the bipartite graph codes for which local purification is possible
using the techniques of~\cite{duer:qc2003a}. In this case, the error
in the destination qubits from preparation is propagated forward to
the next teleportation step. An upper bound on the error rate that the
code needs to handle is given by $e_t=2d_p+d_m+d_b$. If one computes
the error rate exactly, taking into consideration canceling errors,
one gets
\begin{equation}
1-e_t = \begin{array}[t]{cll}
       &(1-d_p)(1-d_m)(1-d_p)(1-d_b) & \textrm{(no error anywhere)}\\
       + &d_p(d_m/3)(1-d_p)(1-d_b) & \textrm{(storage error cancels preparation error}\\&&\textrm{on original qubit)}\\
       + &(1-d_p)(1-d_m)d_p(d_b/3) & \textrm{(Bell-measurement error compensates}\\&&\textrm{for preparation error in} \\ &&\textrm{newly prepared qubit)}\\
       + & (d_p(1-d_m/3)+(1-d_p)d_m)(1-d_p)(d_b/3) & \textrm{(Bell-measurement error compensates}\\&&\textrm{for total error on the original qubit)}
    \end{array}
\end{equation}
If the supremum of the error rates for which the family of codes in
question has arbitrarily small error probability for the encoded qubit
is $e_{\textrm{\tiny max}}$, then a bound on the threshold boundary is
determined by $e_{\textrm{\tiny max}} = e_t$. The currently best lower
bound on $e_{\textrm{\tiny max}}$ for stabilizer codes is
$e_{\textrm{\tiny max}} \approx 0.19 $~\cite{divincenzo:qc1998c}.
According to this estimate, the region for which scalable computation
is possible given balanced $d_m=d_b$ is shown in Fig.~\ref{fig:depth}.

\ignore{
u=[];
v=[];
for s=[0.02:0.0005:.2];
 for t=[0.02:0.0005:.2];
  et = 1 - (1-t)*(1-s)*(1-t)*(1-s);
  et = et - t*(s/3)*(1-t)*(1-s);
  et = et - (1-t)*(1-s)*t*(s/3);
  et = et - (t*(1-s/3)+(1-t)*s)*(1-t)*(s/3);
  if (et < .19);
    u=[u,s];
    v=[v,t];
  endif;
 endfor;
endfor;
u=[u,0];v=[v,0];u=[u,0];v=[v,0.1];u=[u,0.1015];v=[v,0];
plot(u,v, '.');
gset term postscript
gset output "graphics/gthcurve.ps"
plot(u,v,'.');
}

\begin{herefig}
\label{fig:depth}
\begin{picture}(0,4)(0,-3.7)
\nputgr{0,0}{t}{height=3.5in}{gthregion}
\nputbox{2.28,-3.5}{tl}{$d_m=d_b$}
\nputbox{-2.35,-.08}{tr}{$d_p$}
\end{picture}
\nopagebreak
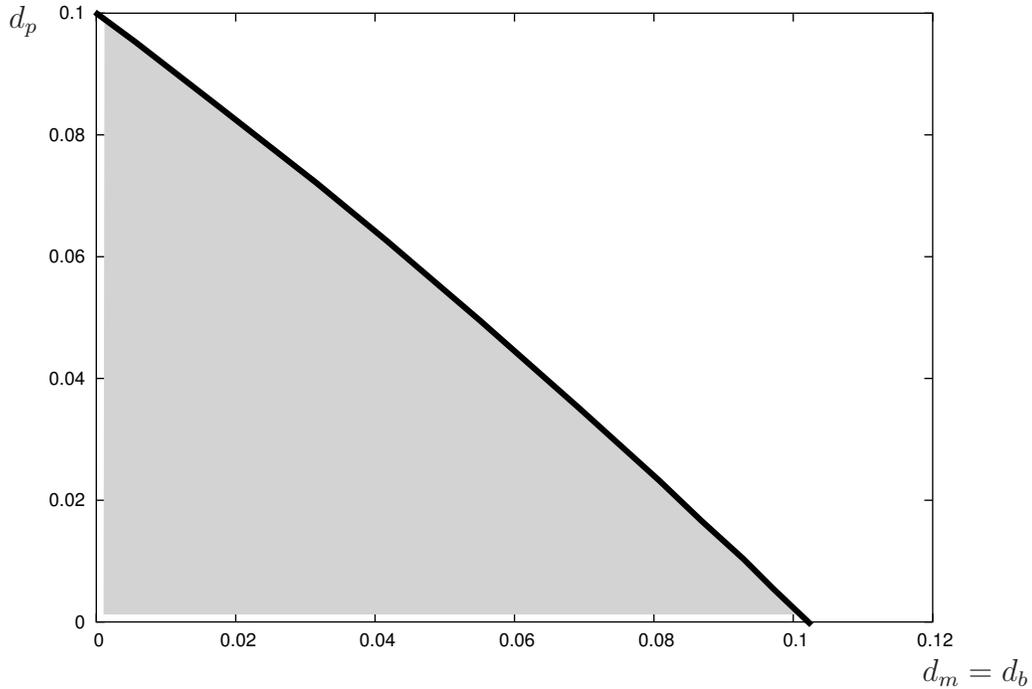
\herefigcap{Region for which scalable computation for 
depolarizing errors is possible. Tolerable error rates are in the gray region,
strictly below the upper boundary. For small memory and Bell-measurement
error rates, preparation errors close to $0.1$ per qubit can
be tolerated. Since preparation errors are likely to dominate
because of the higher number of operations required to deliver
the state, this is the regime that is likely to be relevant.}
\end{herefig}

\section{Discussion}

The work reported here shows that the maximum tolerable error rates
depend strongly on the error model. If the errors are constrained,
then they can be tolerated much better than depolarizing errors.  In
particular, if errors can be detected, tolerable error rates for
computation are above $20\;\%$.  This should
be strong motivation to build in error detection when engineering
quantum devices and designing error-correction strategies.

The use of teleportation demonstrates yet again the now well-known
versatility of this basic quantum communication protocol.  It is worth
noting that frequent use of teleportation in a computation implicitly
solves the leakage problem.  This is the problem where qubits are lost
from the computation without the event being detected, either by
physical loss of the underlying particles, or by the particle's state
leaving the qubit-defining subspace. In every teleportation step, the
destination qubits are fresh, and any previously leaked qubits
contribute only to errors in the Bell measurements.  These errors can
be treated just like other errors.

Further work is required to determine the applicability of this work
to practical quantum computation. For example, the scalability
assumptions include massive parallelism in manipulating quantum bits,
but no communication or classical processing latencies. It may be
possible to reduce the latencies to a constant for relatively slow
quantum computers with measurement times comparable to unitary gate
times, but the necessary architectures that achieve this need to be
further investigated.  Realistically, it is necessary to consider the
actual overheads required for implementing fault-tolerant
computation. In general, experience shows that the overheads grow very
rapidly as the maximum tolerable error rates are approached.  As a
result, it is a good idea to engineer computational devices to have
error-rates that are substantially lower.  In the case of the
techniques used in this paper, the overheads are dominated by the
complexity of state preparation.  Little attempt has been made to
prepare the states more efficiently than by the rejection method: Any
time an error is detected, discard the state and start from
scratch. Since theoretical scalability permits any polynomial
overhead, further optimization is not considered here.  As a result,
overheads can be extremely large. Future investigations
may yield significant improvements in efficiency without large decreases
in tolerated error rates.

\begin{acknowledgments}
Portions of this work were done at Los Alamos National Laboratory.
This work was supported by the Department of Energy (DOE contract
W-7405-ENG-36) and the U.S. National Security Agency. This paper is a
contribution of the National Institute of Standards and Technology, an
agency of the U.S. government, and is not subject to U.S. copyright.
\end{acknowledgments}

\bibliography{journalDefs,qc,thresh}

\end{document}